\documentclass{article}
\usepackage{epsfig}
\usepackage{float}
\usepackage{amsmath}

\newcommand{\be}{\begin{equation}}
\newcommand{\ee}{\end{equation}}

\begin{document}

\title{Synchronization in fiber lasers arrays}

\author{A. ACEVES$^{1}$ and J.-G. CAPUTO$^{2}$ }

\maketitle

{\normalsize \noindent
$^1$: Department of Mathematics, \\
Southern Methodist University,\\
Dallas, Texas, USA\\
E-mail: aaceves@smu.edu ,\\
$^2$:Department of Mathematics,\\
University of Arizona,\\
Tucson, AZ, 85719, USA,\\
E-mail: caputo@insa-rouen.fr\\
}

\date{\ }

\begin{abstract}
We consider an array of fiber lasers coupled through the
nearest neighbors. The model is a generalized nonlinear
Schroedinger equation where the usual Laplacian is replaced
by the graph Laplacian. For a graph with no symmetries, 
we show that there is no resonant transfer of energy between the
different eigenmodes. We illustrate this and confirm our
result on a simple graph. This shows that arrays of fiber ring lasers 
can be made temporally coherent.
\end{abstract}

\section{Introduction}
The dynamics of coupled oscillators continues
to be an active area of research due in large measure to its ubiquitous 
presence in a wide range of disciplines such as  social, life and 
neuro-sciences \cite{nature}. 
It also applies in the traditional areas of classical and 
quantum mechanics, in particular in nonlinear optical systems describing
phenomena such as light localization in waveguide arrays.

This work concentrates on a simple model to understand how
synchronization can be achieved when light is propagating in
ideally identical optical fibers coupled by a suitable scheme.
Two well known models 
encapsule most of the behavior considered here: \noindent
(i) The Kuramoto (KM) model
\begin{equation}
{d\phi_i\over dt}={\tilde w_i} +K \sum_j sin(\phi_i-\phi_j)
\end{equation}
\noindent
(ii) The discrete nonlinear Schroedinger (DNLS) model
\begin{equation} \label{dnls}
i{du_i\over dt} = c(u_{i+1}+ + u_{i-1}) + |u_i|^2 u_i
\end{equation}
where the latter is the most relevant to our study. 
The collective (synchronous) behavior of these coupled optical fibers 
would produce a maximum power electromagnetic output. 
Historically, research in optical fiber technology has been driven
by the increased demand for better communication systems and a natural 
by-product has been the development of fiber lasers \cite{richardson}. 
These are interesting for applications like for example
material cutting where they are starting to replace solid state devices.
To increase their power output, several groups are considering
arrays of fiber lasers.
Here light is amplified in individual, decoupled fiber amplifiers 
which are then coupled within a single cavity or in a second cavity. 
The grand challenge is to achieve a coherent power output that
scales with the square of number of elements in the array. 
In this context, temporal coherence is the equivalent of 
synchronization of oscillators in an array.
In practice, this synchronization has 
proven to be very difficult. Typically the efficiency
dramatically diminishes as the number
of fiber amplifiers increases. Amongst the various combining schemes
a most interesting case is that of Fridman et.al \cite{fridman}.
There the authors combine passively as many as 25
fiber lasers in a two dimensional array. They analyze the power output
in a short time interval over many round trips and observe two main effects.
First, even though the efficiency of the
phase locking is around 20 to 30\%, there are rare events where it
exceeds 90\%. The second observation is that the
efficiency depends on the coupling architecture, the best performance
is when connectivity is increased.

These fiber arrays at high intensities are well described
by the DNLS equation (\ref{dnls}).
A number of studies of this model suggest that nonlinearity
can enhance coherence. Following this, here we propose a more general
coupling scheme where the usual discrete Laplacian operator in
the DNLS equation is replaced by a graph Laplacian. 
Our main result is that when choosing a specific form of the coupling
we are able to "separate" the modes so the nonlinearity
will only couple them weakly. Then the system can be considered
as temporally coherent.
This holds for any graph such that the eigenvalues
of the Laplacian are distinct, i.e. a graph with no symmetries.
We illustrate this by analyzing a simple graph and confirm this
result numerically. We calculate the coherence factor of 
the array for different initial mode distribution and show that 
fiber ring lasers in an array can be synchronized.\\
The article is organized as follows.
We introduce the model in section 2 and show that 
the nonlinearity does not couple the linear eigenmodes on
average so that their average amplitude is constant. In section 3, we 
illustrate this general result on a particular graph. This system is
analyzed numerically in section 4. There we confirm our predictions
and consider different mode distributions. Conclusions are
presented  in section 5.

\section{The model }

\begin{figure}
\centerline{ 
\epsfig{file=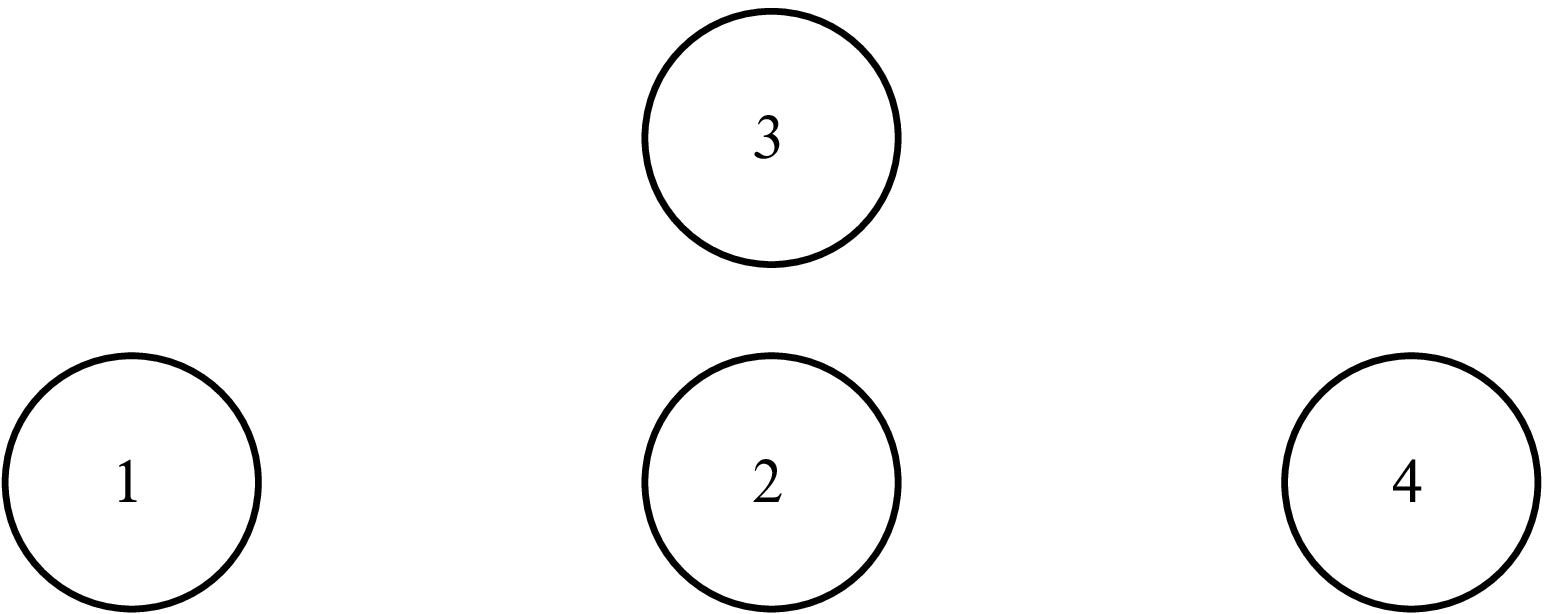,height=2 cm,width=4 cm,angle=0}
\hskip 1.5cm
\epsfig{file=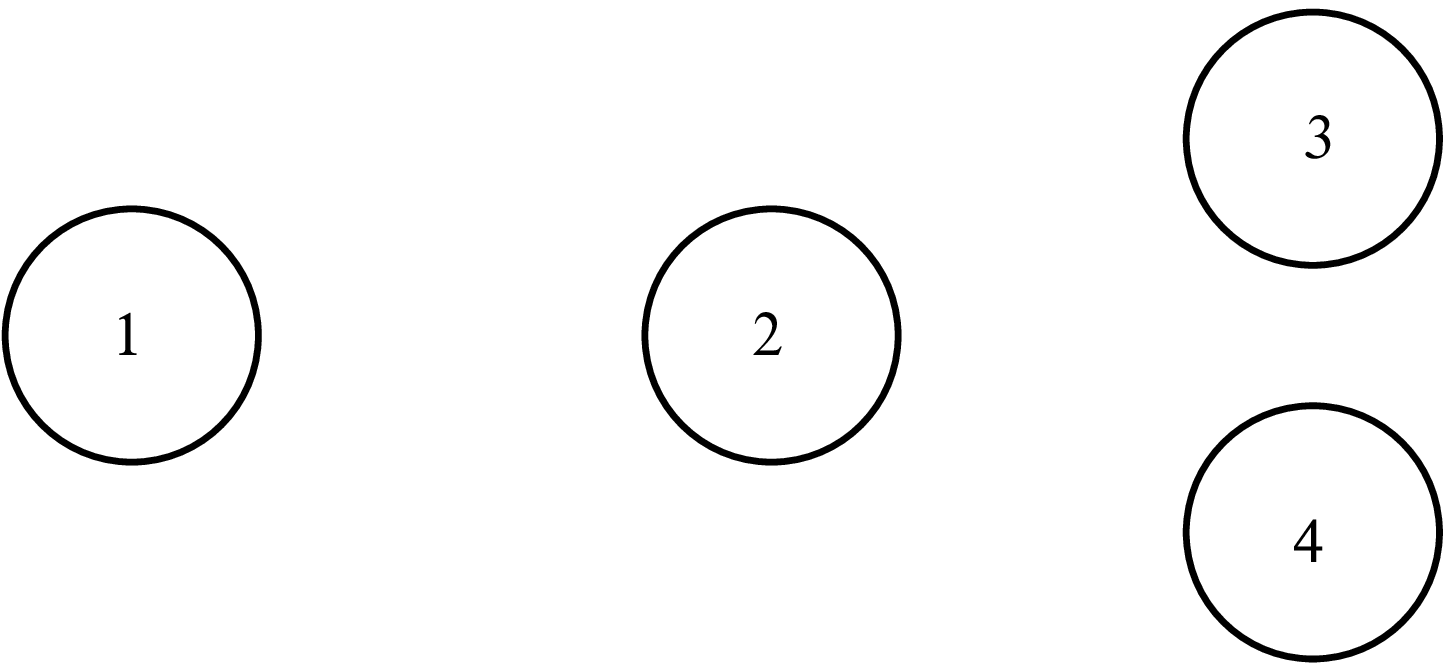,height=2 cm,width=4 cm,angle=0} }
\vskip .5cm
\centerline{ \epsfig{file=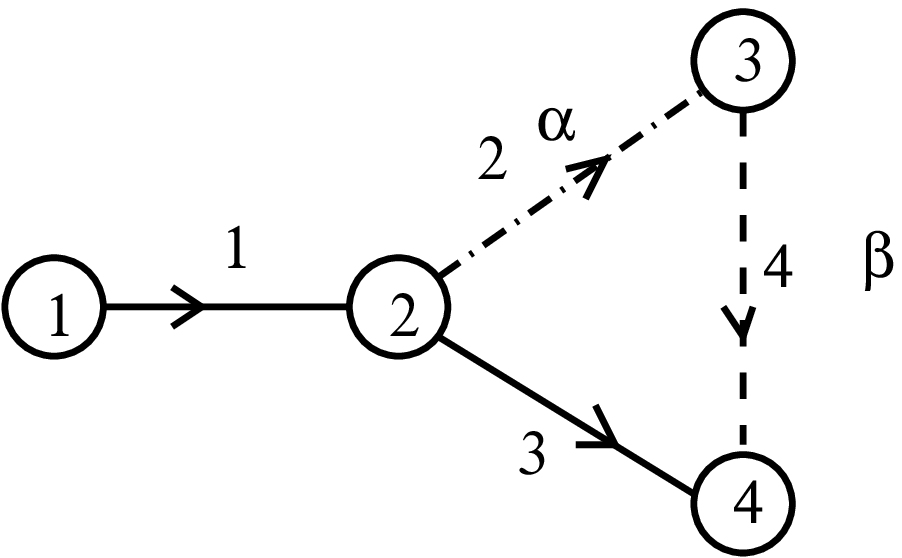,width=0.5\linewidth,angle=0}}
\caption{The top panels presents two physical arrangements of 
a four fiber array. On the left
the fibers 2 and 3 are closer and thus coupled stronger
than the other pairs of fibers. On the top right panel 
the couple 3 and 4 are closer.
The bottom panel shows the graph used to model these devices. 
The arrows on the
branches are oriented arbitrarily and the coupling parameters $\alpha,\beta$
are adjusted to
represent each of the physical configurations.}
\label{f1}
\end{figure}

We propose a design of a fiber array where the coupling is purposely made
irregular. 
We expect that in a network the coupling will
be different due to the arrangement of the single fibers
in relation to others. 
Fig. \ref{f1}. presents the cross-section of the distribution of fibers 
in the array
and the associated network or graph used to model it.
A consequence of this is that we need to 
generalize the classical  2D nonlinear Schroedinger
equation 
$$ iu_t = \Delta u + |u|^2 u~,$$
to the graph NLS
\be\label{gnls} 
 iu_t = G u + |u|^2 u,
\ee
where the standard 2D Laplacian is replaced by the graph Laplacian
$G$. This operator is a natural extension of the continuous Laplacian
see \cite{cks11} for examples of its application.
The specific form of (\ref{gnls}) written for the graph shown in Fig. \ref{f1}
gives
$$
\left\{
\begin{array}{llll}
i\dot{u_1}=\left(u_2-u_1\right) + |u_1|^2 u_1 ~,\\
i\dot{u_2}=-\left(u_2-u_1\right) +\alpha \left(u_3-u_2\right)
+\left(u_4-u_2\right) +|u_2|^2 u_2 ~,\\
i\dot{u_3}=-\alpha \left(u_3-u_2\right) 
+ \beta\left(u_3-u_4\right)+|u_3|^2 u_3~, \\
i\dot{u_4}=\left(u_2-u_4\right) -\beta\left(u_3-u_4\right) + |u_4|^2 u_4,
\end{array}
\right.
$$
where $u_i, ~i=1-4$ are the values of the field amplitude, where 
$\beta$ is the coupling for the branch between nodes
1 and 2, where $\alpha$ is the coupling for the branch between nodes 2 and
3 and where 1 is the coupling for the branch between nodes 2 and 4.
The system of equations above can be written symbolically as
\be\label{gnls2} 
iU_t = G U + N(U).\ee
In the above equation the vector $U$ , the graph Laplacian $G$ and the
nonlinearity are respectively
$$U=\left(
\begin{array}{l}
u_1\\
u_2\\
u_3\\
u4\end{array}
\right),~~
G = \left(
\begin{array}{cccc}
-1 & 1 & 0 & 0 \\
1 & -2-\alpha & \alpha & 1 \\
0 & \alpha & -\alpha-\beta & \beta\\
0 & 1 &  \beta & -1-\beta \\
\end{array}
\right)
, ~~
N(U)=\left(
\begin{array}{l}
|u_1|^2 ~u_1\\
|u_2|^2 ~u_2\\
|u_3|^2 ~u_3\\
|u_4|^2 ~u_4\end{array}
\right).$$
This system preserves the "mass"
$$ M= \sum_i |u_i|^2, $$
and the energy
$$ H=  \sum (\nabla u) ^2 + \sum_i |u_i|^2-|u_i|^4, $$
where $\nabla$ is the discrete gradient associated to the graph \cite{cks11}.

Since the matrix $G$ is symmetric it is natural, following \cite{cks11}, 
to use as a basis for $U$ the eigenvectors of $G$, such that
$G z_i = -\omega_i^2 z_i$.  These are the
columns of the orthogonal $Z$ matrix such that
\be\label{eig_g}
G Z = Z D,\ee
where $D$ is the diagonal matrix of diagonal $-\omega_1^2, -\omega_2^2 ...$.
We then introduce the vector $\Gamma$ of components $\gamma_k, k=1,\dots n$ such that
$$ U = Z \Gamma .$$
In terms of these new coordinates $\gamma_k$, equation (\ref{gnls2}) reduces to
\be\label{dgama1}
i\dot{\gamma_k}=-\omega_k^2 \gamma_k + \sum_{j=1}^4 z_{jk} |u_j|^2 u_j,\ee
where we have used the orthogonality of the eigenvectors of $G$, $Z^{-1}=Z^T$.
The term $|u_j|^2 u_j$ can be written as
$$|u_j|^2 u_j = 
\sum_{lmn} z_{jl}z_{jm}z_{jn} \gamma_l \gamma_m \gamma_n^*~.$$
We then get the final equation
\be\label{dgama}
i\dot{\gamma_k}=-\omega_k^2 \gamma_k 
+ \sum_{jlmn} z_{jl}z_{jm}z_{jn} \gamma_l \gamma_m \gamma_n^*~ .\ee
This is the equation that we will analyze throughout the article.

Let us assume that all the eigenvalues $\omega_j$ are simple.
This is the generic case for a graph without symmetries.
Then we can simplify the
equation (\ref{dgama}) by eliminating the first term on the right hand
side. For that we introduce 
$$\gamma_k = e^{i \omega_k^2 t}\beta_k,$$
to obtain
\be\label{dbeta}
i\dot{\beta_k}=
 \sum_{jlmn} z_{jl}z_{jm}z_{jn} \beta_l \beta_m \beta_n^*~
e^{i (\omega_l^2 +\omega_m^2-\omega_n^2-\omega_k^2)t}~~ .\ee
Only the resonant terms such that
\be\label{reson}\omega_l^2 +\omega_m^2-\omega_n^2-\omega_k^2=0,\ee
will contribute on the long term to $\dot{\beta_k}$. 
Then, because the frequencies are all different, the resonant
condition is satisfied if $l=n$ and $m=k$. We then obtain the
resonant evolution of $\dot{\beta_k}$ as
\be\label{dbetar}
i\dot{\beta_k}=
 \beta_k \sum_{jl} z_{jl}^2~z_{jk} |\beta_l|^2 
.\ee
This equation is such that the intensity in each mode is constant
$$I_k = |\beta_k|^2,~~\dot{I_k}=0~.$$ 
The solution of equation (\ref{dbetar}) is then
\be\label{beta}
{\beta_k}(t)= \beta_k(0){\rm exp}(-i~t~ \Omega_k),\ee
where the nonlinear correction to the frequency is
\be\label{Omega}
\Omega_k=\sum_{jl} z_{jl}^2~z_{jk} |\beta_l|^2
.\ee
Returning back to the original variables $\gamma_k$ we get
our main result
\be\label{gammat}
{\gamma_k}(t)= \gamma_k(0){\rm exp} \left [ i~t~ (\omega_k^2 -\Omega_k) \right ] .\ee
There is no long term energy transfer between the modes. 
For a given initial condition, the modal distribution is fixed
and the system will keep this for all times.

Before presenting a case study, a few remarks are in place. 
Our analysis is valid for
any graph, of arbitrarily large size. For our approximate
analysis to be valid it is important that eigenvalues be
different and well separated. Then the phase resonance
condition will be well satisfied. We will illustrate this below
by comparing our prediction to the numerical solution of the 
full problem. Also note that
when there are multiple eigenvalues, the change of
variable from $\gamma$ to $\beta$ is no longer valid and the
degeneracy will induce a chaotic temporal dynamics.

\section{A case study: a graph with a swivel}

We now illustrate the above considerations on a simple
example derived from the network shown in Fig. \ref{f1}.
This will enable us to put numbers on the frequencies 
$\omega_k$ and $\Omega_k$. 
In the study \cite{cks11}, we introduced the notion of "soft node"
where the eigenvector has a zero coordinate. These are important
because if applied to them the damping or forcing of the network is
ineffective. We will consider the simplest such network that has
a soft node. It corresponds to the tree with $\beta=0$.

The eigenvalues and eigenvectors $z_i$ (columns of the $z$ matrix)  can be computed analytically, they
are given in Table 1 in terms of $\alpha$, 
$$\delta = \sqrt{4\alpha^2-4\alpha+9},$$
$$n_3= \sqrt{2+(\delta-2\alpha-1)^2/4+(\delta-2\alpha +3)^2/4},$$
and
$$ n_4= \sqrt{2+ (\delta + 2\alpha +1)^2/4+(\delta + 2\alpha -3 )^2/4}.$$
\begin{table} \label{tab2}
\begin{tabular}
{|l | c | c | c | r |}
  \hline
index $i $           &  1    &  2 &  3   &   4 \\ \hline
$\lambda_i= -\omega_i^2$   &  0    & -1 & ${1\over 2}(\delta-2\alpha-3)$ &$-{1\over 2}(\delta +2\alpha +3)$\\
                           &       &    &                                &                                 \\\hline
                           &  1/2    & $1/\sqrt{2}$   &  $1/n_3$                            &  $1/n_4$                             \\
$z_i$                      &  1/2    & 0   & ${1\over 2n_3}(\delta-2\alpha-1)$ & $ -{1\over 2n_4}(\delta + 2\alpha +1)$ \\
                           &  1/2    & 0   & $-{1\over 2n_3}(\delta-2\alpha +3)$& $ {1\over 2n_4}(\delta + 2\alpha -3 )$\\
                           &  1/2    & $-1/\sqrt{2}$  &  $1/n_3$                            &   $1/n_4$                            \\ \hline
 \end{tabular}
\caption{Eigenmodes $(\lambda_i, z_i)$ for the tree.}
\end{table}

\begin{figure} [H]
\centerline{\epsfig{file=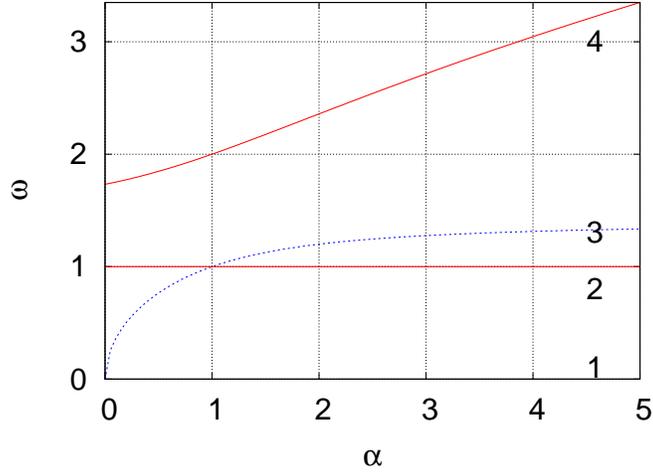,width=0.8\linewidth,angle=0}}
\caption{ Plot of the eigenfrequencies $\omega_i,~i=1-4$ as a function of
$\alpha$ for the graph shown in Fig. \ref{f1}} with $\beta=1$.
\label{f1c}
\end{figure}
\begin{figure} [H]
\centerline{
\epsfig{file=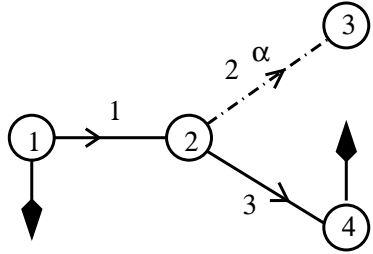,width=0.4\linewidth,angle=0}}
\caption{Schematic representation of
the constant mode $z_2$, corresponding to  $\omega_2=1$}.
\label{f3a}
\end{figure}

The mode $z_2$ which is independent of $\alpha$ is shown schematically
in Fig. \ref{f3a} where we have plotted the magnitude and sign
of the coordinate using vertical arrows. The arrows are opposite and equal
for $z_{2}^1$ and $z_{2}^4$ because the nodes 1 and 4 play a symmetric
role i.e. the graph is invariant by the automorphism transforming node
1 to node 4 \cite{crs01}.
\begin{figure} [H]
\centerline{
\epsfig{file=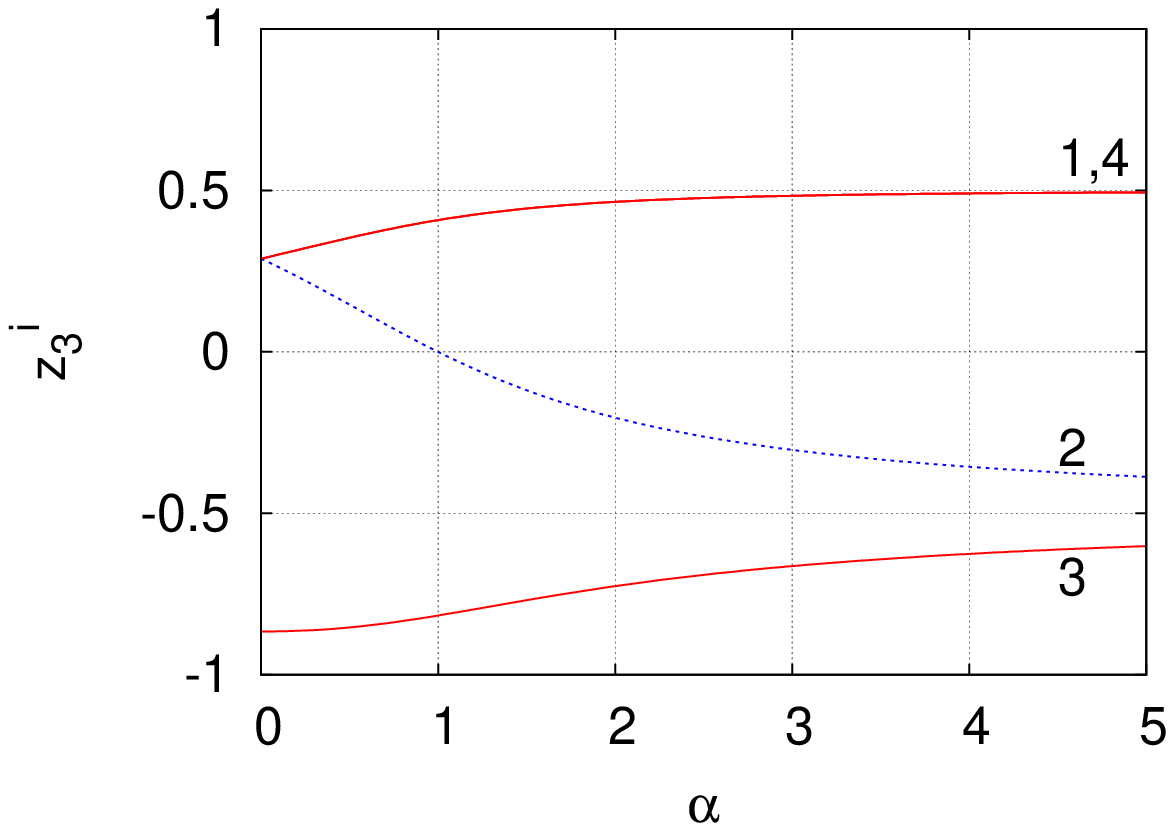,width=0.5\linewidth,angle=0}
\epsfig{file=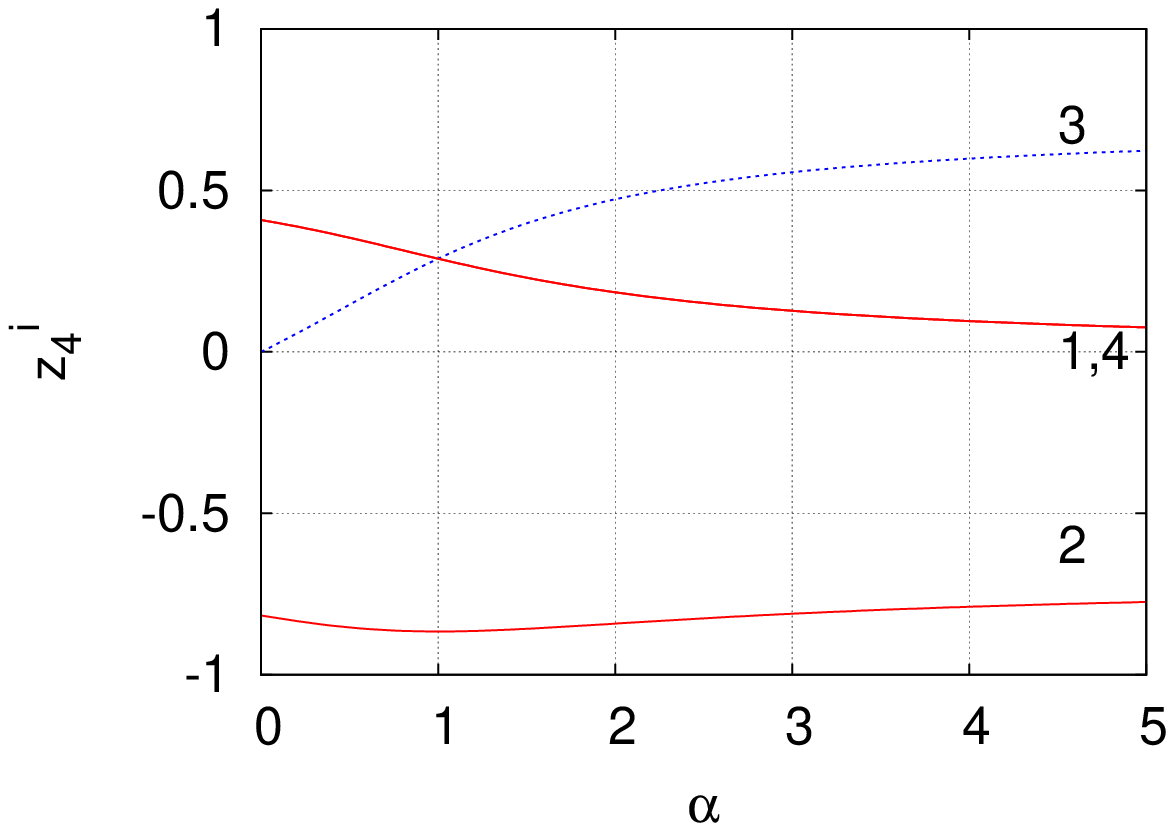,width=0.5\linewidth,angle=0}
}
\caption{Plot of the components of the non trivial normalized eigenvectors $z_3$
(left panel) and $z_4$ (right panel)
as a function of $\alpha$.}
\label{f2}
\end{figure}

Let us compute the frequencies of oscillations $\Omega_k$ of the resonant modes.
The right hand side of relation (\ref{beta}) can be written in matrix form
$$M I = \Omega$$
where the vector $I= \left ( \begin{matrix}
 I_1 \cr
 I_2 \cr
 I_3 \cr
I_4  
\end{matrix}  \right)$  with $I_j=|\beta_j|^2$ , where the vector $\Omega=\left ( \begin{matrix}
\Omega_1  \cr
\Omega_2  \cr
\Omega_3  \cr
\Omega_4  
\end{matrix}  \right)$ and where the matrix $M$ has a general term
$$m_{kl}=\sum_{j} z_{jk}~z_{jl}^2 .$$
Introducing the matrix $Z$ from Table 1 and computing $M$ using
Matlab we get the general linear system 
\be\label{sysio}
\left ( \begin{matrix}
 0.5 & 0.5 & 0.5 & 0.5 \cr
 0 & 0 & 0 & 0 \cr
 0 & m_{32} & m_{33} & m_{34} \cr
 0 & m_{42} & m_{43} & m_{44} 
\end{matrix}  \right )
\left ( \begin{matrix}
 I_1 \cr
 I_2 \cr
 I_3 \cr
I_4  
\end{matrix}  \right) 
=
\left ( \begin{matrix}
\Omega_1  \cr
\Omega_2  \cr
\Omega_3  \cr
\Omega_4
\end{matrix}  \right).
\ee
This matrix enables to compute the nonlinear corrections $\Omega_k$ 
once the mode distribution $\gamma_k$ is given.
The matrix $M$ has rank 3 and its image is
$$Im(M)= \{ {\vec x},{\vec z}, {\vec t} \}.$$
Note that $\Omega_2=0$.
Taking for example $I_1=2, ~I_2=I_3=I_4=0$ we get
$\Omega_1=0$ and $\Omega_2=\Omega_3=\Omega_4=0$.


%
%
\section{Numerical results}

To validate our approach we now solve numerically the original 
equations (\ref{gnls2})
for the tree configuration described above ($\beta=0$). We chose
the initial mode configuration and let it evolve. The resolution was done
using the ode45 subroutine of Matlab \cite{matlab}.

\begin{figure} [H]
\centerline{
\epsfig{file=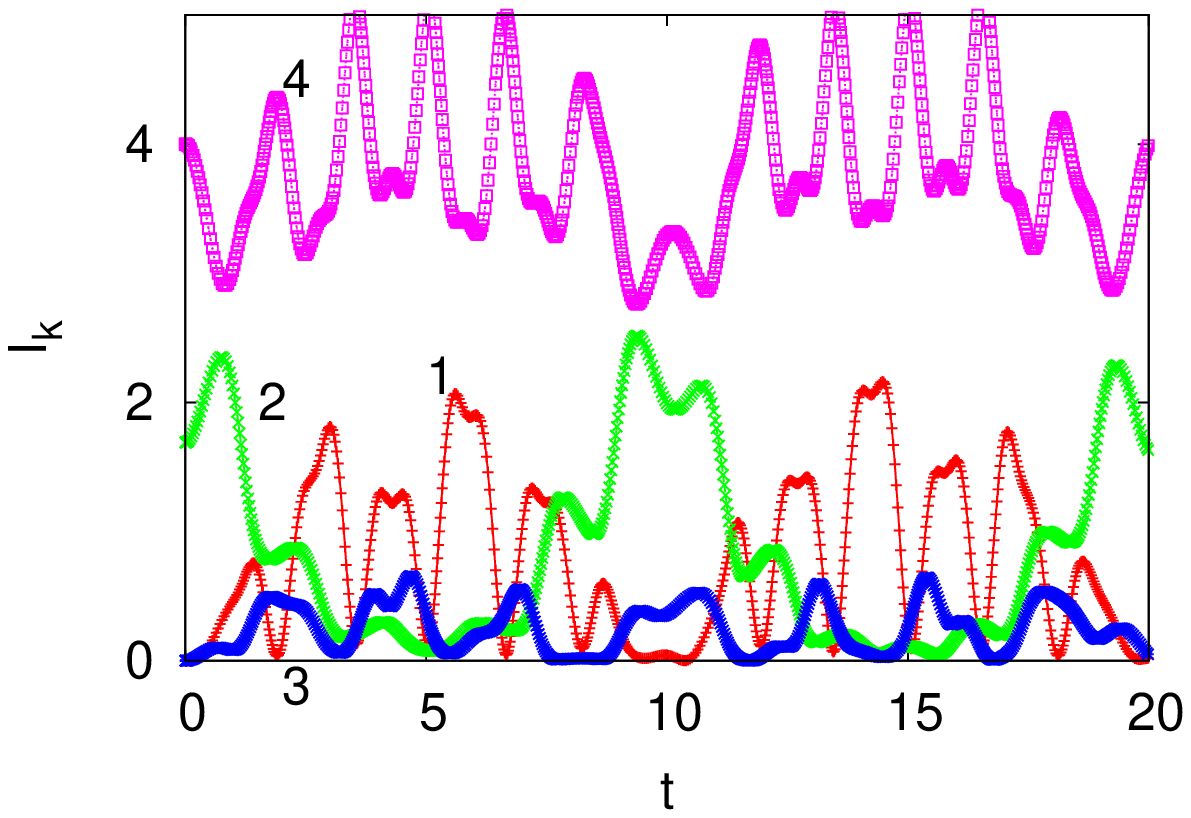,width=0.5\linewidth,angle=0}
\epsfig{file=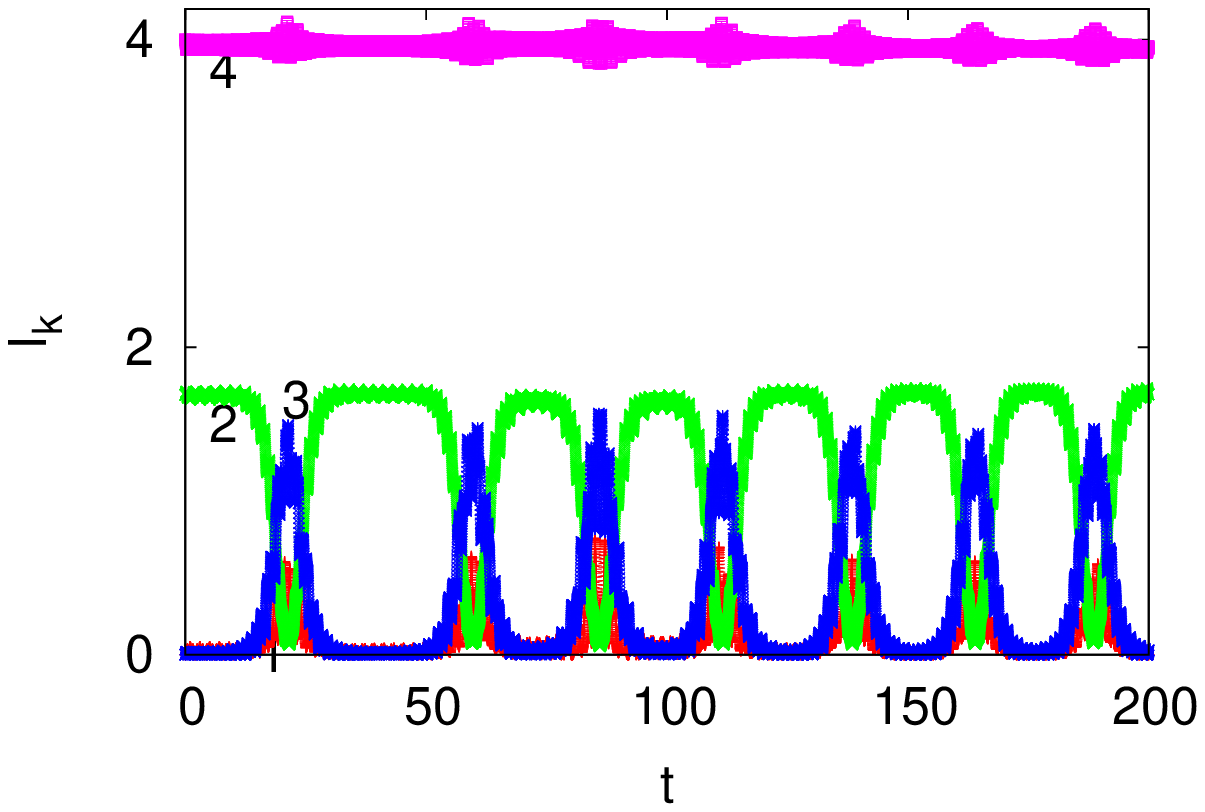,width=0.5\linewidth,angle=0}
}
\caption{Time evolution of the energies in each mode $I_k$ for $k=1-4$
for two values of the coupling, $\alpha=0.5$ (left panel)  and $\alpha=3.5$ (right panel).}
\label{f4}
\end{figure}

Fig. \ref{f4} shows the time evolution of the energies $I_k$ for $k=1-4$
for two values of the coupling, $\alpha=0.5$ (left panel)  and $\alpha=3.5$ (right panel).
The initial conditions are the same $\beta_1=0,\beta_2=1.3,\beta_3=0$ and $\beta_4=2$.
Notice how $I_4$ is approximately constant in agreement with our prediction.
On the other hand in both cases the modes 2 and 3 have close frequencies so they couple strongly.
For $\alpha=0.5$ the frequencies are closer so that more interaction occurs.
For $\alpha=3.5$ we have $\omega_4 >> \omega_2$ so the phase term in  (\ref{reson}) is
larger and the approximation is better. Furthermore, it is clear that for
this case a slow dynamics will show a fixed point for mode 4; and homoclinic
orbits for modes 1-3. All this is seen in the right panel of 
Fig. \ref{f4}.

To relate our  results to figures of merit in experimental work on 
coherent beam combining, and in particular paying attention to that 
of Fridman et al. 
\cite{fridman}, we introduce the coherence factor $C$
\be\label{coh}
C = {|\sum_i u_i| \over \sum_i |u_i|}~.\ee
For the modes $\omega_i,~i=2-4$ this coherence factor will be smaller
than one because the eigenvectors $z_i$ have to have positive and negative
coordinates, which could be interpreted as being in (same sign) or  out 
(different sign) of phase data. As an example consider that 
$$\gamma_1=\gamma_2=\gamma_3=0, ~~ \gamma_4\neq 0  ~.$$ 
Then for $\alpha$ large we have
$$u_2 =z_{24} \gamma_4\approx -0.8 \gamma_4,
~~u_3=z_{34} \gamma_4\approx 0.6 \gamma_4 ~~,$$
and $u_1=u_4=0$. This clearly represents an out of phase combination
leading to a low coherence coefficient value,
$$C = {|u_2+u_4| \over |u_2|+|u_4|}= 0.14 .$$
If however one considers instead of the nodes $u$, the modes $\gamma$,
say by applying the linear transformation given by the matrix $Z$,
then the coherence coefficient becomes 1. This could be done using
transformational optics.

\begin{figure} [H]
\centerline{
\epsfig{file=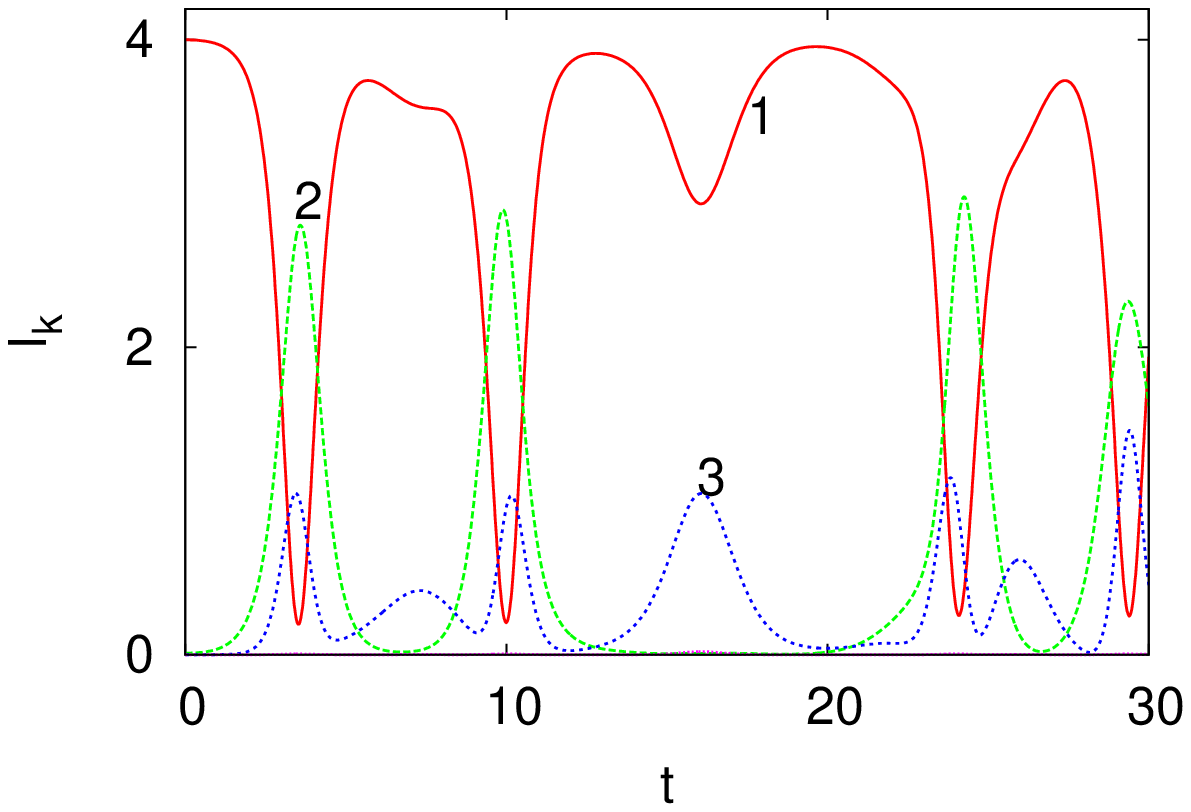,width=0.5\linewidth,angle=0}
\epsfig{file=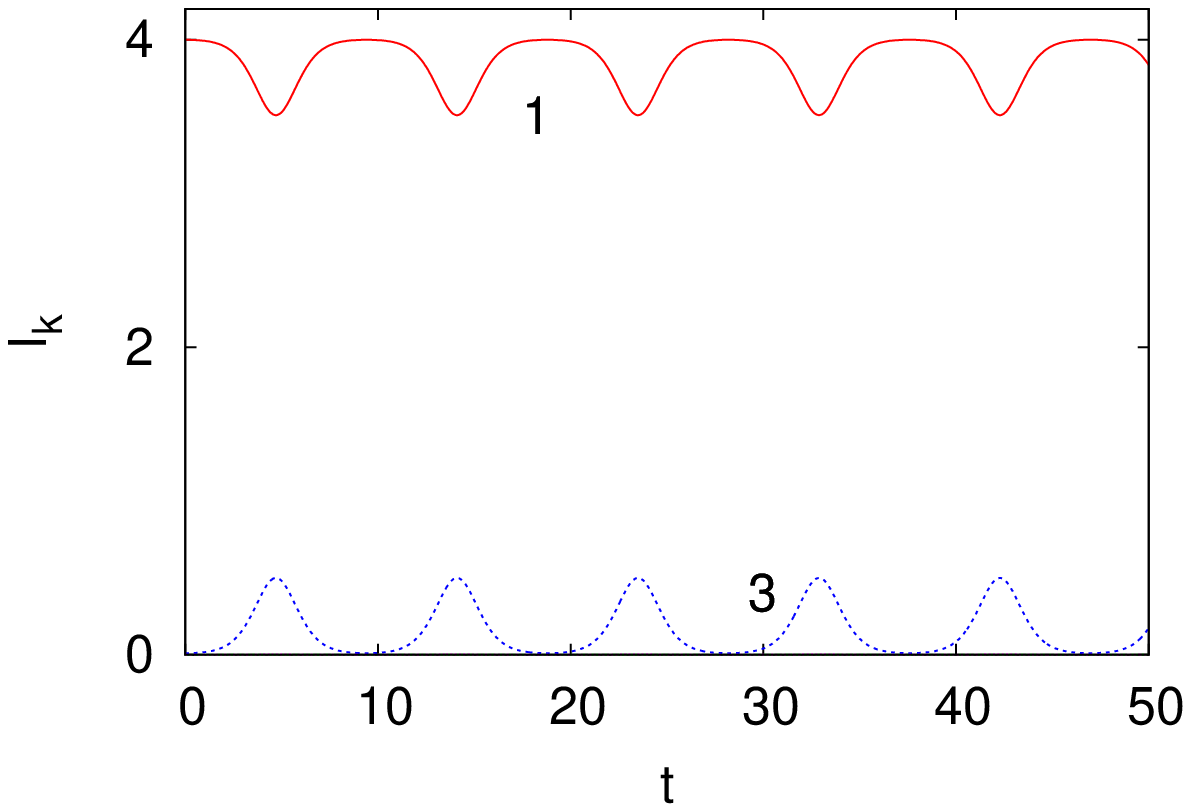,width=0.5\linewidth,angle=0}
}
\caption{
Time evolution of the energies in each mode $I_k$ for $k=1-4$
for $\gamma_1=2,~\gamma_4=0,$ . The left panel is for
$\gamma_2=0.1,~ \gamma_3=0$ and the right panel for
$\gamma_3=0.1,~ \gamma_2=0$.
The other parameter is $\alpha=3.5$
}
\label{f5}
\end{figure}

Another direction would be to use the Goldstone mode $z_1$
corresponding to all node components $u_i$ equal. For this
we need to guarantee that the mixing with the other frequencies
remains small. This can be done for the tree as we now show.
We fix $\gamma_1=2,~\gamma_4=0,$~ and choose two sets of the 
pair $(\gamma_2,\gamma_3), $ $\gamma_2=0.1,~ \gamma_3=0$
and $\gamma_2=0,~\gamma_3=0.1$ . The results are shown in
Fig. \ref{f5} with $\gamma_2=0.1,~ \gamma_3=0$ on the left panel
and $\gamma_3=0.1,~ \gamma_2=0$ on the right panel. As expected 
on the left panel we see a stronger coupling than on the right panel 
because
$$\omega_1^2 - \omega_2^2 < \omega_1^2 - \omega_3^2 ,$$
so that the rotating wave approximation leading to (\ref{gammat})
is better satisfied. Considering the coherence factor,
the parameters $\gamma_1=2,~\gamma_4=0,$ and $\gamma_3=0.1,~ \gamma_2=0$ 
will give $C\approx 1$.
The results of this section show that if the array is "prepared" in a given
mode, it will stay on that mode on average. For this preparation, one
can force the array at resonance like in \cite{cks11}.
As an example, for our tree configuration we select mode 4 by damping 
nodes 1 or 4 and forcing only nodes 2 or 3.

\section{Conclusions}

Motivated by studies in fiber laser arrays and how the 
different coupling schemes between fibers can affect the
coherence property of the output, we
introduced an array where the coupling between fibers
is irregular. This leads to a discrete nonlinear Schroedinger
equation where the usual discrete Laplacian is replaced by
a graph Laplacian.

To connect our work with experiments, we computed a coherence 
coefficient and showed that we can synchronize the array.
This study then opens the way of using a new approach based in graph 
Laplacians for designing new coupling schemes for arrays of fiber lasers.

\section{Acknowledgements}

The work of A. A. has benefited from regular collaborations with Dr. Erik 
Bochove from the US AFRL.
J.G. C. acknowledges partial support 
from the "Grand Reseau de Recherche, Transport Logistique et Information"
of the Haute-Normandie region in France.
The authors acknowledge the Centre de Ressources Informatiques de Haute
Normandie where most of the calculations were done.
J. G. C. is on leave from Laboratoire de Math\'ematiques, INSA de
Rouen, France.

\end{document}